\documentclass[12pt,aps,prb,preprint]{revtex4}

\usepackage{amsmath}
\usepackage{graphicx}   

\def\d{{\rm d}}

\def\vf{{\varphi}}
\def\be{\begin{equation}}
\def\ee{\end{equation}}
\def\bea{\begin{eqnarray}}
\def\eea{\end{eqnarray}}
\begin{document}

\title{slipping  and rolling on an inclined plane}

\author{Cina Aghamohammadi$^{a)}$, \& Amir Aghamohammadi$^{b)}$}\email{mohamadi@alzahra.ac.ir}
 \affiliation{$^{a)}$Department of Electrical Engineering, Sharif University of Technology, P.O. Box. 11365-11155 , Tehran, Iran}

  \affiliation{$^{b)}$Department of Physics, Alzahra University, Tehran 19938-91176, Iran}

\date{\today}

\begin{abstract}
    In the first part of the article using a direct calculation two-dimensional motion of a particle sliding on an inclined plane is investigated for general values of friction coefficient ($\mu$). A parametric equation for the trajectory of the particle  is also obtained. In the second part of the article the motion of a sphere on the inclined plane is studied. It is shown that the evolution equation for the contact point of a sliding sphere is similar to that of a point particle sliding on an inclined plane whose friction coefficient is $\frac{7}{2}\ \mu$.
    If $\mu> \frac{2}{7}\tan\theta$, for any arbitrary initial velocity and angular velocity the sphere will roll on the inclined plane after some finite time. In other cases, it will slip on the inclined plane.
    In the case of rolling center of the sphere moves on a parabola. Finally the velocity and angular velocity of the sphere are exactly computed.
\end{abstract}

\maketitle

\section{Introduction}
One of the standard problems in elementary mechanics is a particle sliding on an inclined plane. Usually it is assumed that the motion is one-dimensional \cite{Hall},\cite{KLep}. However there are also some textbooks\cite{Irodov},\cite{GHR}in which  two-dimensional motion of a particle sliding on an inclined plane for special choice of friction coefficient is considered. In a recent article\cite{SL} it is shown that there is an analogy between curvilinear motion on an inclined plane and the pursuit problem.

In this article using a direct calculation two-dimensional motion of a particle on an inclined plane of angle $\theta$  is studied. In section II sliding of a particle  on an inclined plane is investigated for general values of friction coefficient, $\mu$. We obtain particle's velocity in terms of $\vf$, the slope of particle's trajectory. At first general behavior of the particle's velocity at large times has been considered.
It is shown that for $\mu>\tan \theta$ after a finite time the particle's velocity will vanish, and at this time $\vf$ is also equal to zero. For $\mu\geq 1$, at large times the particle moves in a straight line.
An exact calculation is also done and a relation between time, $t$, and $\vf$ is obtained.
In section III, the motion of a sphere on the inclined plane is studied. Depending on the friction coefficient and initial velocity and angular velocity of the sphere, it may roll or slide on the inclined plane.
The evolution equation for the contact point of a sliding sphere is obtained and it is shown that it is similar to
that of a point particle sliding on an inclined plane whose friction coefficient is $\frac{7}{2}\ \mu$. For $\mu> \frac{2}{7}\tan\theta$ depending on initial velocity, sphere may initially slip but it will roll after some finite time. For $\mu\leq \frac{2}{7}\tan\theta$ it will slip forever. It is shown that when the sphere rolls on an inclined plane generally the center of sphere moves on a parabola. Finally the velocity and angular velocity of the sphere are exactly computed.

\section{sliding a particle on an inclined plane}
\begin{figure}
\begin{center}
\includegraphics{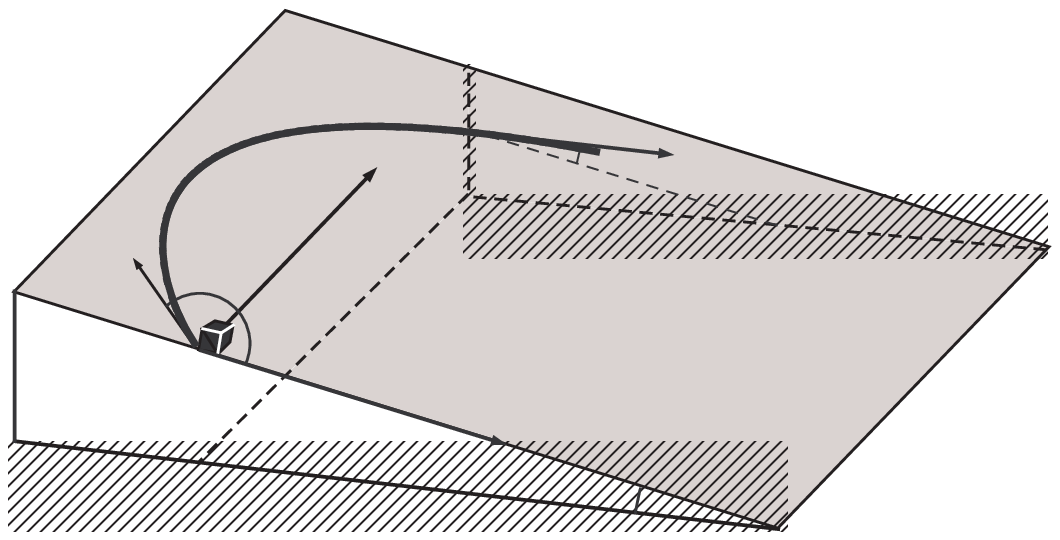}
\setlength{\unitlength}{1mm}
\put(-50,6){$\theta$} \put(-101,29.6){${\bf v}_0$}\put(-75.5,40.5){$y$} \put(-49,40.7){$\varphi$}\put(-62,9.5){$x$}
\put(-41,41){${\bf e}_v$\put(-48.8,-12.7){$\vf_0$}}
\caption{\label{fig:inclinedplane}}{A particle sliding on an inclined plane }
\end{center}
\end{figure}

Let's consider an inclined plane of angle $\theta$. We want to study sliding a particle of mass $m$ on this plane.  Here it is assumed that friction coefficient,$\mu$, is constant and the problem is solved for general values of $\mu$.  Newton's equation of motion is
\be\label{iplane00}
m\ddot {\bf r}= mg\sin\theta\ {\bf i}-\mu m g \cos\theta \ {\bf e}_v,
\ee
where ${\bf e}_v$ is the unit vector in the direction of particle's velocity.
See Fig. (\ref{fig:inclinedplane}).
Let's define $\lambda:=\mu \cot \theta$, where $\vf$ is the slope of particle's trajectory at the point $(x,y)$. Then equation (\ref{iplane00}) recasts to
\bea\label{iplane01}
m\ddot x&=& mg\sin\theta(1-\lambda\cos\vf) ,\cr
m\ddot y&=&-\lambda mg\sin\theta\sin\vf
\eea
Newton's equation of motion of the particle tangential to its trajectory is
\be\label{iplane01-2}
m\ddot s=mg\sin\theta(\cos\vf-\lambda),
\ee
where $s$ is the length parameter along the particle's trajectory. Let's solve the problem for different values of $\lambda$. Note that  $\lambda$ is a nonnegative parameter.

\subsection{$\lambda=0$}
This case corresponds to a frictionless plane. The particle has a constant acceleration $g\sin \theta$, and  the component of velocity along $y$ axis, $v_y$, remains constant. Then the trajectory of the particle is generally a parabola.
\subsection{$\lambda=1$}
This case has been partially studied in Refs. 3, and 4. Setting $\lambda=1$, (\ref{iplane01}) and (\ref{iplane01-2})
recast to
\bea\label{iplane01-3}
m\ddot x&=& mg\sin\theta(1-\cos\vf),\cr
m\ddot y&=&- mg\sin\theta\sin\vf,\cr
m\ddot s&=&mg\sin\theta(\cos\vf-1)
\eea
it is seen that
\be\label{iplane02}
\ddot s+\ddot x=0,
\ee
Then $\dot s+\dot x$ should be constant. Assuming initial velocity to be ${\bf v}_0$ (see fig. \ref{fig:inclinedplane}), then
\be
 \dot s+\dot x=v_0(1+\cos\vf_0).
\ee
Then using
\bea\label{iplane04}
&&\dot x= \dot s\cos \vf,\cr
&&\dot y= \dot s\cos \vf,
\eea
one arrives at
\bea\label{iplane05}
&&\dot s= \frac{v_0(1+\cos\vf_0)}{1+\cos \vf},\cr&&\cr
&&\dot x= \frac{v_0\cos \vf(1+\cos\vf_0)}{1+\cos \vf},\cr&&\cr
&&\dot y= \frac{v_0\sin \vf (1+\cos\vf_0)}{1+\cos \vf}.
\eea
As $\vf$ is the slope of particle's trajectory at the point $(x,y)$, then
$\tan \vf=\displaystyle{\frac{\d y}{\d x}=\frac{\dot y}{\dot x}}$, and
\be\label{iplane06}
\dot \vf (1+\tan^2 \vf)=\frac{\ddot y\dot x-\ddot x\dot y}{\dot x^2}
\ee
Using (\ref{iplane01-3}) and (\ref{iplane06}),  $\dot\vf$ can be obtained
\be\label{iplane07}
\dot \vf= -\frac{g\sin \theta\ \sin \vf(1+\cos \vf)}{v_0(1+\cos\vf_0)}.
\ee
It is seen that for any  $0<\vf<\pi$, $\dot \vf$ is negative. So $\vf $ is a decreasing function of time. Let's consider its behavior at large times,
\be\label{iplane07-2}
\dot \vf\approx -\frac{2g\sin \theta}{v_0(1+\cos\vf_0)}\ \vf,\quad \Rightarrow \quad \vf \propto
 {\rm e}^{-\displaystyle{\frac{2gt\sin \theta }{v_0(1+\cos\vf_0)}}}.
\ee
At large times $\vf $ goes to zero, the particle's trajectory is a straight line and it's velocity will be
\bea\label{iplane08}
&&\lim_{t\to \infty }\dot s=\lim_{t\to \infty }\dot x =\frac{v_0(1+\cos\vf_0)}{2},\cr
&&\lim_{t\to \infty }\dot y= 0.
\eea
There is a maximum value for y,
\bea
\int_0^{y_{\rm max}}\d y&=&\frac{v_0^2(1+\cos\vf_0)^2}{g\sin \theta}\int_0^{\vf_0}\frac{{\d \vf}}{(1+\cos \vf)^2}\cr&&\cr
&=&\frac{v_0^2(1+\cos\vf_0)^2}{4g\sin \theta}\int_0^{\vf_0}\d \vf\left( (1+\tan^2\frac{\vf}{2})
+ \tan^2\frac{\vf}{2}(1+\tan^2\frac{\vf}{2})\right)\cr&&\cr
y_{\rm max}&=& \frac{v_0^2\sin\vf_0}{g\sin \theta}(1+\frac{1}{3}\tan^2\frac{\vf_0}{2}),
\eea
\subsection{$\lambda\ne 1$}
Combining (\ref{iplane01}) and (\ref{iplane01-2})
gives
\be
\ddot x+ \lambda \ddot s=g\sin \theta (1-\lambda^2),
\ee
or
\be
\dot x+ \lambda \dot s=g\sin \theta (1-\lambda^2)t +v_0(\lambda+\cos\vf_0),
\ee
where we have used of boundary condition. Similar to the preceding case, one may obtain
\bea\label{iplane11}
&&\dot s=  \frac{g\sin \theta (1-\lambda^2)t +v_0(\lambda+\cos\vf_0)}{\lambda+\cos \vf},\cr&&\cr
&&\dot x= \frac{\left(g\sin \theta (1-\lambda^2)t +v_0(\lambda+\cos\vf_0)\right)\cos \vf}{\lambda+\cos \vf},\cr&&\cr
&&\dot y= \frac{\left(g\sin \theta (1-\lambda^2)t +v_0(\lambda+\cos\vf_0)\right)\sin \vf}{\lambda+\cos \vf},
\eea
from which we obtain
\be\label{iplane12}
\dot \vf= -\frac{g\sin \theta(\lambda +\cos \vf)\sin \vf}{g\sin \theta (1-\lambda^2)t +v_0(\lambda+\cos\vf_0)}.
\ee
\subsubsection{$0<\lambda<1$}
For special choice of initial conditions the particle's velocity  may become zero but the friction is not large enough to keep it at rest. At large times $\vf $ goes to zero. Let's consider its behavior at large times, or small $\vf$'s
\be\label{iplane13}
\dot \vf\approx  -\frac{\vf}{(1-\lambda)t},\quad \Rightarrow \quad \vf
\propto t^{\displaystyle{-\frac{1}{1-\lambda}}}.
\ee
In both cases ($\lambda=1$, and $\lambda<1$) at large times $\vf \to 0$. In the previous case it approaches to zero exponentially, and in the latter case  in the form of power law.
However in both cases at large times the particle's trajectory is a straight line. So, at large times for $\mu<\tan \theta$, the particle goes down the inclined plane with a constant acceleration.

\bea\label{iplane14}
&&\lim_{t\to \infty }\dot s\sim g\sin \theta (1-\lambda)t, \cr &&
\lim_{t\to \infty }\dot x \sim g\sin \theta (1-\lambda)t,\cr
&&\lim_{t\to \infty }\dot y= 0,
\eea
\subsubsection{$\lambda>1$}
In this case friction coefficient is larger than previous cases and  the particle will be finally at rest. It is seen from (\ref{iplane11}), that $\dot s$ will be zero at the time $T$,
\be\label{iplane12-3}
T=\frac{v_0(\lambda+\cos\vf_0)}{g\sin \theta (\lambda^2-1)}.
\ee
When the particle's velocity vanishes because of friction it will remain at rest.
At the time $t=T-\epsilon$
\be\label{iplane12-2}
\dot \vf= -\epsilon^{-1}\big[\frac{g\sin \theta(\lambda +\cos \vf)\sin \vf}{g\sin \theta (\lambda^2-1)}\big]\Big\vert_{t=T-\epsilon}.
\ee
So $\vf$ decreases rapidly until it reaches zero, and when the particle's velocity approaches zero its velocity is in the $x$ direction.
\subsection{exact solution}
We studied large time behavior of the particle's motion for different cases.
Now let's do an exact calculation. using (\ref{iplane12}), one may arrive at
\be\label{iplane15}
\frac{-(1-\lambda^2)\sin \vf\ \d \vf}{(\lambda +\cos \vf)(1-\cos^2\vf)}
= \frac{g\sin \theta(1-\lambda^2)\d t}{g\sin \theta (1-\lambda^2)t +v_0(\lambda+\cos\vf_0)},
\ee
which can be written as
\bea\label{iplane16}
\int_{\cos\vf_0}^{\cos\vf}\d \cos \vf'\big[\frac{1}{\lambda +\cos \vf'}+\frac{(1-\lambda)}{2(1-\cos \vf')}
-\frac{(1+\lambda)}{2(1+\cos \vf')}
\big]&=&\cr &&\cr \int_0^t\frac{g\sin \theta(1-\lambda^2)\d t'}{g\sin \theta (1-\lambda^2)t' +v_0(\lambda+\cos\vf_0)}.&&
\eea
Integrations can be done easily and gives
\be\label{iplane17}
t= \frac{v_0(\lambda+\cos\vf_0)}{g\sin\theta (1-\lambda^2)}
\left\{\frac{(\lambda +\cos\vf)\sin \vf_0}{(\lambda +\cos\vf_0)\sin \vf}\cdot\displaystyle{
\left(\frac{\tan ({\vf}/{2})}{\tan ({\vf_0}/{2})}\right)^\lambda}-1\right\}
\ee
In the limiting case $\lambda=1$ changes to
\be\label{iplane17-2}
t= \frac{v_0(1+\cos\vf_0)}{g\sin\theta}
\left\{ \ln\left( \frac{\tan (\vf/2)}{\tan(\vf_0/2)}\right)
+\frac{1}{1+\cos \vf}-\frac{1}{1+\cos \vf_0}\right\}
\ee
In fig. 2, $\vf$ is drawn in terms of time $t$, for two values of $\lambda $, and five different value of $\vf_0$.
For $\lambda=2$, $\vf $ goes to zero rapidly, but for $\lambda=0.5$ it goes to zero asymptotically.

Using (\ref{iplane11}) and (\ref{iplane17}), one can obtain velocity components
\bea\label{iplane17-3}
&&\dot x= v_0\sin \vf_0 \cot \vf\left( \frac{\tan(\vf/2)}{\tan(\vf_0/2)}\right)^{\lambda}
,\cr&&\cr
&&\dot y= v_0\sin \vf_0 \left( \frac{\tan(\vf/2)}{\tan(\vf_0/2)}\right)^{\lambda},
\eea

\begin{figure}
\begin{center}
\includegraphics{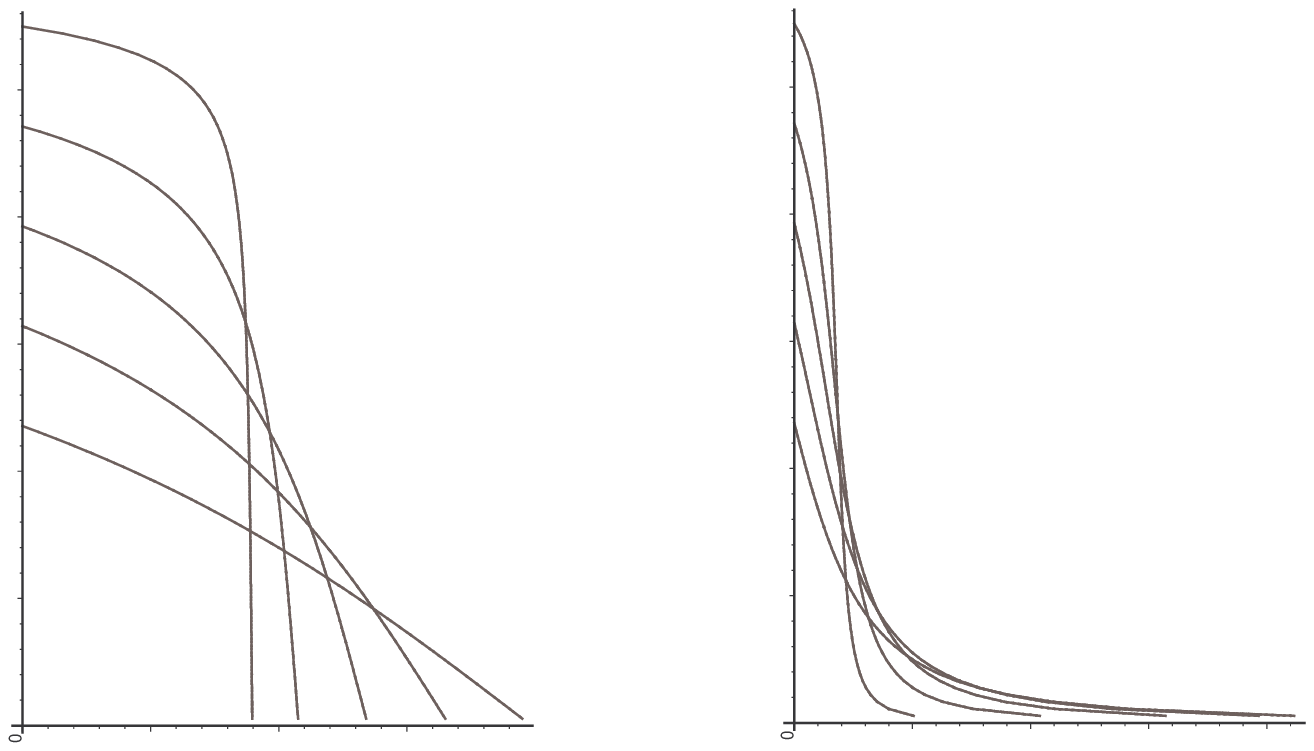}
\setlength{\unitlength}{1mm}
\put(-134.2,80.6){$\vf$} \put(-55.8,80.6){$\vf$} \put(-87.7,1){$gt\sin \theta/v_0$}\put(-9,1){$gt\sin \theta /v_0$}
\put(-143.3,74.2){$7\pi/8$}\put(-143.3,64.2){$3\pi/4$}\put(-143.3,54.2){$5\pi/8$}\put(-143.3,44.2){$\pi/2$}
\put(-143.3,34.2){$3\pi/8$}
\put(-64.8,74.2){$7\pi/8$}\put(-64.8,64.2){$3\pi/4$}\put(-64.8,54.2){$5\pi/8$}\put(-64.8,44.2){$\pi/2$}
\put(-64.8,34.2){$3\pi/8$}
\put(-123.2,1.2){$0.2$}\put(-110.2,1.2){$0.4$}\put(-97.2,1.2){$0.6$}\put(-110.2,-3.6){a)}\put(-33.2,-3.6){b)}
\put(-44,1.2){$2$}\put(-31.8,1.2){$4$}\put(-19.7,1.2){$6$}
\put(-109.2,65){$\lambda=2$} \put(-30.8,65){$\lambda=0.5$}
\caption{\label{fig:inclinedplane2}}{$\vf$ in terms of time $t$ for five different values of $\vf_0$,  a) for $\lambda=2$, $\vf $ goes to zero rapidly, b) for $\lambda=0.5$ it goes to zero asymptotically}
\end{center}
\end{figure}

Now a parametric equation for the trajectory of the particle can be obtained. It is easy to obtain
\bea
\frac{{\rm d}x}{{\rm d}\vf}&=&-\frac{v_0^2\sin^2\vf_0}{g\sin \theta(\tan ({\vf_0}/{2}))^{2\lambda}} \frac{(\tan ({\vf}/{2}))^{2\lambda}\cot\vf}{\sin^2 \vf},\cr
&&\cr
\frac{{\rm d}y}{{\rm d}\vf}&=&-\frac{v_0^2\sin^2\vf_0}{g\sin \theta(\tan ({\vf_0}/{2}))^{2\lambda}} \frac{(\tan ({\vf}/{2}))^{2\lambda}}{\sin^2 \vf},
\eea
which can be integrated and leads to a parametric equation for the trajectory of the particle,
\bea
x&=&-\frac{v_0^2\sin^2\vf_0}{4g\sin \theta(\tan ({\vf_0}/{2}))^{2\lambda}}\left\{\left[ \frac{(\tan ({\vf}/{2})))^{2\lambda-2}}{2\lambda-2}- \frac{(\tan({\vf}/{2}))^{2\lambda+2}}{2\lambda+2}\right]-\left[\vf\to\vf_0 \right]\right\}\cr
&&\cr
y&=&-\frac{v_0^2\sin^2\vf_0}{2g\sin \theta(\tan ({\vf_0}/{2}))^{2\lambda}}\left\{\left[ \frac{(\tan ({\vf}/{2}))^{2\lambda-1}}{2\lambda-1}+ \frac{(\tan ({\vf}/{2}))^{2\lambda+1}}{2\lambda+1}\right]-\left[\vf\to\vf_0 \right]\right\}.
\eea
The equation of the trajectory of the particle in the case $\lambda=1$ is
\bea
x&=&\frac{v_0^2(1+\cos\vf_0)^2}{16g\sin \theta}\left[\tan^4\frac{\vf}{2}-\tan^4\frac{\vf_0}{2}-4\ln\left(\frac{\tan({\vf}/{2})}{\tan(({\vf_0}/{2})}\right) \right]\cr
&&\cr
y&=&\frac{v_0^2(1+\cos\vf_0)^2}{6g\sin \theta}\left[3\tan\frac{\vf_0}{2}+\tan^3\frac{\vf_0}{2}-3\tan\frac{\vf}{2}-\tan^3\frac{\vf}{2} \right].
\eea
\begin{figure}
\begin{center}
\includegraphics{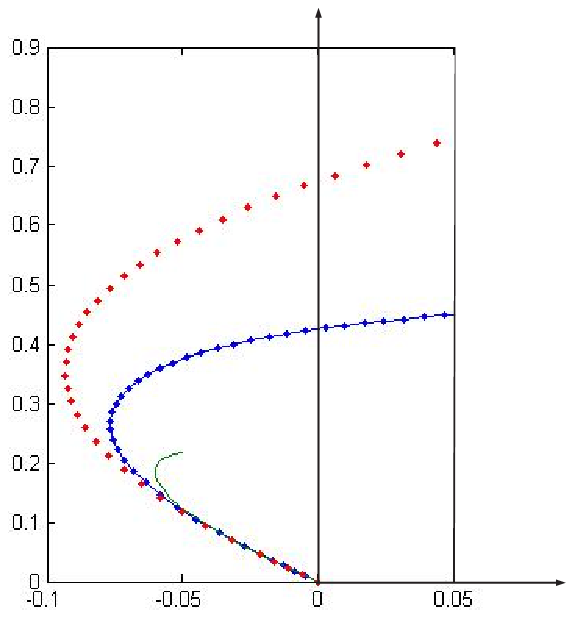}
\setlength{\unitlength}{1mm}
\put(-65.2,23){$\lambda=2$} \put(-38.5,55){$\lambda=0.3$}\put(-38.5,37){$\lambda=1$}\put(-26.5,10){$x$}\put(-53.8,70.6){$y$}
\caption{\label{fig:inclinedplane01-2}}{Trajectory of three projectile with the same  velocity ${\bf v}_0=-{\bf i}+2{\bf j}$\,(m/s) on an inclined plane of angle $\theta=\pi/3$, from time $t=0$ s till $0.5$ s for different value of $\lambda$}
\end{center}
\end{figure}
In fig. 3 trajectory of three projectile with the same  velocity is drawn from time $t=0$ s till $0.5$ s for different value of $\lambda$. It is based on numerical calculation. All three have the same initial velocity ${\bf v}_0=-{\bf i}+2{\bf j}$\,(m/s), and the angle of inclined plane is $\theta=\pi/3$.
\section{The motion of a sphere on an inclined plane}
In this section we want to study the motion of a sphere with the radius $R$, and mass
$m$ on an inclined plane. Depending on friction coefficient and initial velocity and angular velocity
the sphere it may roll or slide on the inclined plane.
Newton's equation of motion for the sphere is
\bea\label{iplane18}
&&m\ddot {\bf r}_{\rm cm}= mg\sin\theta\, {\bf i}+{\bf f}\cr
&&I\dot {\mathbf\Omega}={\bf R}\times {\bf f},
\eea
where ${\bf R}=-R\,{\bf k}$, and $I=2mR^2/5$ is the moment of inertia of the sphere with respect to its center, ${\bf r}_{\rm cm}$ is radius vector from the origin to the center of mass, and ${\mathbf \Omega}$ is the angular velocity of the sphere.
Let's first consider the rolling of sphere.

\subsection{rolling a sphere on an inclined plane}
 Rolling constraint demands the velocity of contact point of the sphere with the inclined plane, $A$, to be zero.
Then
\be\label{iplane19}
{\bf v}_A=\dot{\bf r}_{\rm cm}+\mathbf\Omega\times {\bf R}=0.
\ee
Differentiating the above equation with respect to time and using (\ref{iplane18}), one obtains
\be\label{iplane20}
{\bf f}=-\frac{I}{R^2}\, \ddot {\bf r}_{\rm cm}
\ee
and
\bea\label{iplane21}
\ddot{\bf r}_{\rm cm}&=&\frac{5g\sin \theta}{7}{\bf i},\cr
{\bf f}&=&-\frac{2mg\sin \theta}{7}{\bf i}.
\eea
The sphere rolls on the inclined plane if $f\leq \mu mg\cos\theta$. Then the  rolling
occurs if $\mu\geq \frac{2}{7}\tan\theta$. If sphere rolls on the inclined plane, friction will be a constant force. Then for arbitrary initial velocity and angular velocity, trajectory of sphere's center is generally a parabola.

\subsection{sliding a sphere on an inclined plane}
If the sphere slips then the velocity of contact point $A$ is not zero, and friction is
\bea\label{iplane22}
{\bf f}&=&- \mu mg \cos \theta {\bf e}_A\cr
&=&- \mu mg \cos \theta (\cos \vf {\bf i}+\sin \vf {\bf j})
\eea
where ${\bf e}_A=\displaystyle{\frac{{\bf v}_A}{v_A}}$ is the unit vector along velocity of contact point, and $\vf$ is the angle between ${\bf v}_A$ and ${\bf i}$.
 Using (\ref{iplane18}) time evolution equation of ${\bf v}_A$ is
\be\label{iplane23}
m \dot {\bf v}_A=mg\sin \theta {\bf i}-\frac{7}{2}\ \mu mg\cos \theta{\bf e}_A.
\ee
As it is seen the evolution equation for the velocity of contact point of a sliding sphere with friction coefficient $\mu$ is exactly the same  evolution equation of velocity of a point particle sliding on an inclined plane whose friction coefficient is $\frac{7}{2}\ \mu$.  Compare (\ref{iplane23}) with (\ref{iplane00}).
So it does not need to solve the equation for ${\bf v}_A$, and  all the previous results can be used only by replacing $\lambda$ with $\lambda_s:= \frac{7}{2}\ \mu\cot \theta$,
e.g. ${\bf v}_A$ can be obtained through (\ref{iplane17-3})
\bea\label{iplane23-2}
&& v_{Ax}=v_{0Ax} \sin \vf_0 \cot \vf\left( \frac{\tan(\vf/2)}{\tan(\vf_0/2)}\right)^{\lambda_s}
,\cr&&\cr
&& v_{Ay}= v_{0Ay}\sin \vf_0 \left( \frac{\tan(\vf/2)}{\tan(\vf_0/2)}\right)^{\lambda_s},
\eea
where $v_{0Ax}$ and $v_{0Ay}$ are components of initial velocity of contact point.
It should be noted that this is not enough  to know  ${\bf v}_{\rm cm}$, and ${\mathbf \Omega}$.

Similar to sliding particle three cases may occur:
\subsubsection{$\lambda_s>1\,(\mu> \frac{2}{7}\tan\theta)$}
if $\lambda_s>1\,(\mu> \frac{2}{7}\tan\theta)$, As it was shown in the previous section after a finite time, $T_r$,
\be
T_r=\frac{v_{A0}(\lambda_s+\cos\vf_0)}{g\sin \theta (\lambda_s^2-1)},
\ee
${\bf v}_A$ will become equal to zero. When the  rolling constraint holds true, the sphere will roll.
At the time $T_r$, friction is along the $x$ direction, and is a constant force. So the sphere rolls down the inclined plane, and its center moves on a parabola.
\subsubsection{$\lambda_s=1\, (\mu= \frac{2}{7}\tan\theta)$}
Using our previous results on sliding particle, at large times  ${\bf v}_A$  approaches to a constant value
\be
{\bf v}_A\to \frac{v_{A0}(1+\cos \vf_0)}{2}\ {\bf i}.
\ee
so sphere slides forever on the plane. At large times friction is along the $x$ direction, and is a constant force. In this case at large times $\ddot{\bf r}_A=0$. Then at large times the acceleration of center of mass is
\be
\dot{\bf v}_{\rm cm}=\frac{5g\sin \theta}{7}{\bf i}.
\ee
\subsubsection{$\lambda_s<1\, (\mu< \frac{2}{7}\tan\theta)$}
This case corresponds to $\lambda<1$, of the previous section.  Sphere slides forever on the plane and at large times
${\bf v}_A$ is along the $x$ direction. So at large times friction is along the $x$ direction, and is a constant force. Then the acceleration of center of mass is
\be
\ddot{\bf r}_{\rm cm}=g\sin \theta(1-\mu \cot \theta){\bf i}.
\ee

\subsection{exact solution}
Let's solve exactly the Newton's equation. Integrating (\ref{iplane18}) gives
\bea\label{iplane30}
&&{\bf v}_{\rm cm}={\bf v}_{0\rm cm} + {\bf i}g\sin\theta\ t  +\frac{1}{m}\int_0^t \d t\ {\bf f}\cr
&&{\mathbf\Omega}= {\mathbf\Omega}_0 +\frac{{\bf R}}{I}\times \int_0^t \d t\  {\bf f}.
\eea
To know  ${\bf v}_{\rm cm}$, and ${\mathbf \Omega}$, ${\bf f}= -\mu mg \cos \theta ( {\bf i}\cos \vf+ {\bf j}\sin \vf)$ should be integrated. Two main integrals should be calculated,
\bea
\int_0^t \d t\ \sin \vf&=& \int_0^t \frac{\d \vf}{\dot \vf }\sin \vf\cr&&\cr
                       &=& -\frac{v_{0A}\sin \vf_0}{g\sin \theta (\tan (\vf_0/2))^\lambda}\int_{\vf_0}^{\vf} \frac{\d \vf \, (\tan (\vf/2))^\lambda}{\sin \vf },\cr &&\cr
                       &=&\frac{v_{0A}\sin \vf_0}{\lambda g\sin \theta}\left[ 1- \left(\frac{\tan (\vf/2)}{\tan (\vf_0/2)}\right)^\lambda\right]
\eea

\bea
\int_0^t \d t\ \cos \vf &=& -\frac{v_{0A}\sin \vf_0}{g\sin \theta (\tan (\vf_0/2))^\lambda}\int_{\vf_0}^{\vf} \frac{\d \vf \, \cos \vf (\tan (\vf/2))^\lambda}{\sin^2 \vf },\cr &&\cr
                       &=& -\frac{v_{0A}\sin \vf_0}{2g\sin \theta(\tan (\vf_0/2))^\lambda}
                      \Big\{ \Big[\frac{(\tan (\vf/2))^{\lambda-1}}{\lambda -1}
                       -\frac{(\tan (\vf/2))^{\lambda+1}}{\lambda +1}\Big]\cr &&\cr
                       &&-\Big[ \vf\to \vf_0 \Big]\Big\}
\eea
Substituting  both integrals ${\bf v}_{\rm cm}$, and ${\mathbf \Omega}$ can be obtained. Let's consider the case ${\lambda_s>1}$ as an example and compute
the velocity and angular velocity of sphere when it starts to roll.  At that time ${\bf v}_A$ and $\vf$ are  equal to zero. It can be easily shown that
\bea
&&\int_0^{T_r} \d t\ \sin \vf= \frac{v_{0A}\sin \vf_0}{\lambda_s g \sin \theta}\cr &&\cr
&&\int_0^{T_r} \d t\ \cos \vf =\frac{v_{A0}(\lambda_s\cos\vf_0 +1)}{g\sin \theta (\lambda_s^2-1)}\cr&&\cr
&&\int_0^{T_r} \d t\ {\bf f}=- m v_{0A}\mu \cot \theta \left[ {\bf i}\,\frac{(\lambda_s\cos\vf_0 +1)}{\lambda_s^2-1}+ {\bf j}\,\frac{\sin \vf_0}{\lambda_s}\right]
\eea
After the time $T_r$ the sphere will roll and it's center moves on a parabola.
 \begin{acknowledgments}
We would like to thank M. Khorrami for useful comments. A. A. was partially supported by the
research council of the Alzahra University.
\end{acknowledgments}

\end{document}